\documentclass[10pt,twocolumn,english,aps,prl,superscriptaddress,showpacs,floatfix]{revtex4-1}
\usepackage[latin9]{inputenc}
\usepackage{babel}
\usepackage{amssymb}
\usepackage{amsmath}
\usepackage{graphicx}

\begin{document}

\title{Unconventional topological phase transitions in helical Shiba chains}

\author{Falko Pientka}
\affiliation{\mbox{Dahlem Center for Complex Quantum Systems and Fachbereich Physik, Freie Universit\"at Berlin, 14195 Berlin, Germany}}

\author{Leonid I.\ Glazman}
\affiliation{Department of Physics, Yale University, New Haven, CT 06520, USA}

\author{Felix von Oppen}
\affiliation{\mbox{Dahlem Center for Complex Quantum Systems and Fachbereich Physik, Freie Universit\"at Berlin, 14195 Berlin, Germany}}

\begin{abstract}
Chains of magnetic impurities placed on a superconducting substrate and forming helical spin order provide a promising venue for realizing a topological superconducting phase. An effective tight-binding description of such helical Shiba chains involves long-range (power-law) hopping and pairing amplitudes which induce an unconventional topological critical point. At the critical point, we find exponentially localized Majorana bound states with a short localization length unrelated to a topological gap. Away from the critical point, this exponential decay develops a power-law tail. Our analytical results have encouraging implications for experiment. 
\end{abstract}


\maketitle

{\em Introduction.---}Currently, there is much excitement about topological superconducting phases \cite{review1,review2}. There has been particular interest in one-dimensional topological superconducting phases with $p$-wave symmetry, engineered in hybrid systems based on conventional $s$-wave superconductors, and their Majorana end states \cite{kitaev2,fu08,lutchyn10,oreg10,alicea11,mourik12,das12}. A promising recent proposal involves a chain of magnetic impurities placed on an $s$-wave superconductor and corresponding experiments are under way \cite{bernevig,nagaosa,loss,braunecker,franz,pientka13,ojanen} (see also \cite{beenakker_magnetic,flensberg,morpurgo}). It is envisioned that the magnetic impurities form a spin helix due to the RKKY interaction and induce Shiba bound states \cite{yu,shiba,rusinov,rmp} in the superconducting host. These Shiba states hybridize to form bands, one each for the positive- and negative-energy Shiba states. Once the hybridization becomes strong enough for these two bands to overlap, the system can enter a topological superconducting phase. 

The formation of Shiba bands can be modeled within a tight-binding Bogoliubov--de Gennes Hamiltonian \cite{pientka13}. The model Hamiltonian closely resembles Kitaev's toy model \cite{kitaev2} of spinless $p$-wave superconductors, with the important distinction that the hopping and the pairing are long range \cite{pientka13}. In fact, Shiba states are known to exhibit a slow $1/r$ decay away from the impurity for $r\ll \xi_0$ which crosses over into an exponential decay for $r\gg \xi_0$. For typical Shiba chains, the coherence length $\xi_0$ of the host superconductor is much larger than the impurity spacing $a$ which is comparable to the lattice spacing of the host superconductor. In numbers, one has $\xi_0/a \sim 10^2-10^3$ making a model with a pure $1/r$ decay of hopping and pairing an excellent starting point. In the context of topological phases, this long-range coupling poses interesting questions. Most importantly, it is usually assumed \cite{hasan} that the boundary modes of topological phases, such as Majorana end states, fall off exponentially into the bulk which seems incompatible with long-range coupling. 

In this paper, we provide an analytical theory for the surprising localization properties of the Majorana end states in helical Shiba chains, with important implications for experiment. A crucial property of helical Shiba chains is that as a consequence of long-range coupling, it displays an unconventional topological critical point as a function of the helix and Fermi wavevectors $k_h$ and $k_F$ \cite{pientka13}. The critical point is located exactly at $k_h = k_F$ in the limit $\xi_0\to\infty$ and remains close to it for finite $\xi_0$. Thus right at or near the critical point, 
the spin helix satisfies the condition for Bragg reflection which induces a strong tendency towards localizing the Majorana end states, competing with the delocalizing tendency of the long-range coupling. This may result in a localization length of the order of a few impurity sites, making isolated Majoranas accessible in experimentally feasible chains containing only a few dozen atoms.

\begin{figure*}[tbp]
\begin{centering}
\includegraphics[width=.32\textwidth]{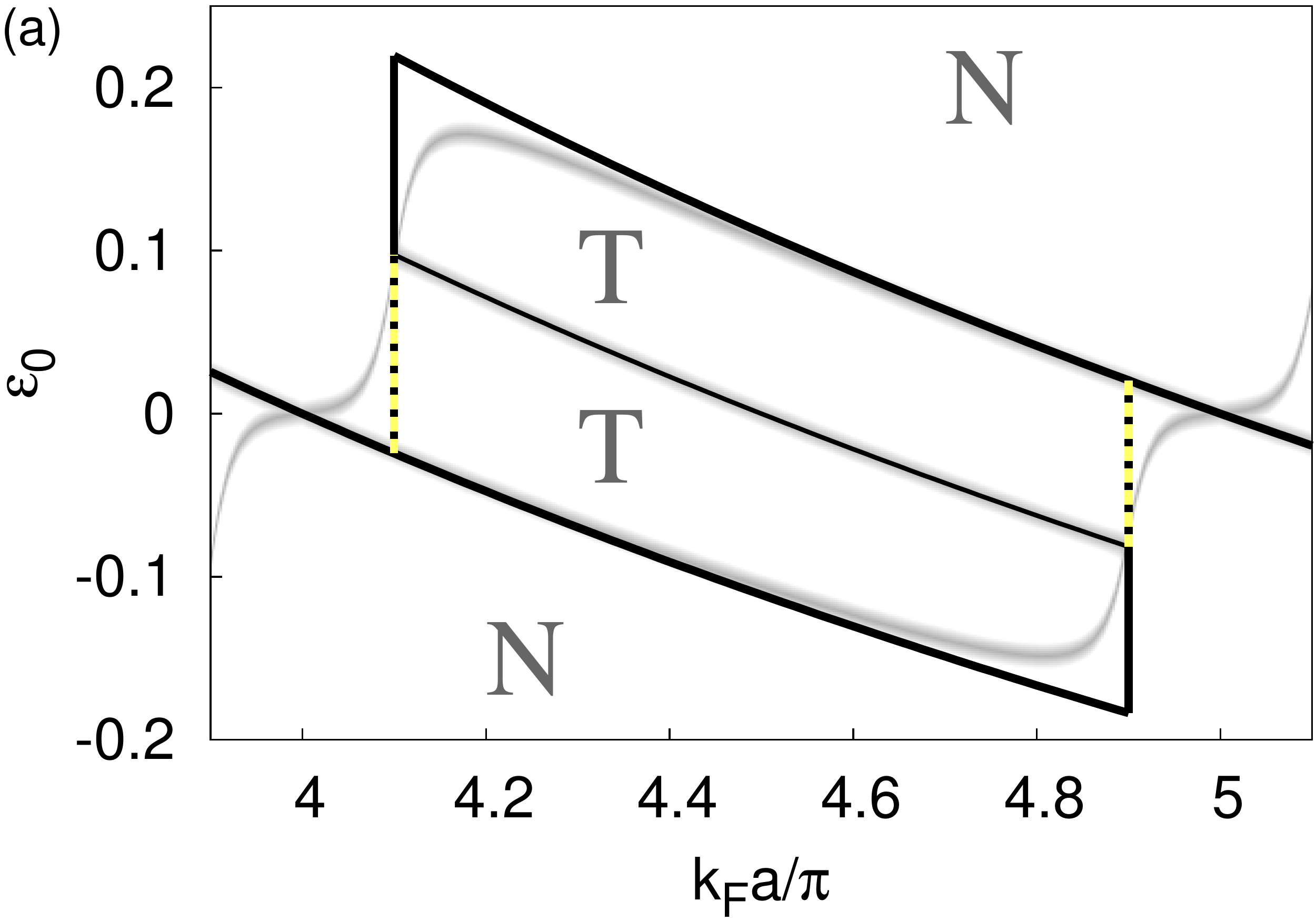}
\includegraphics[width=.32\textwidth]{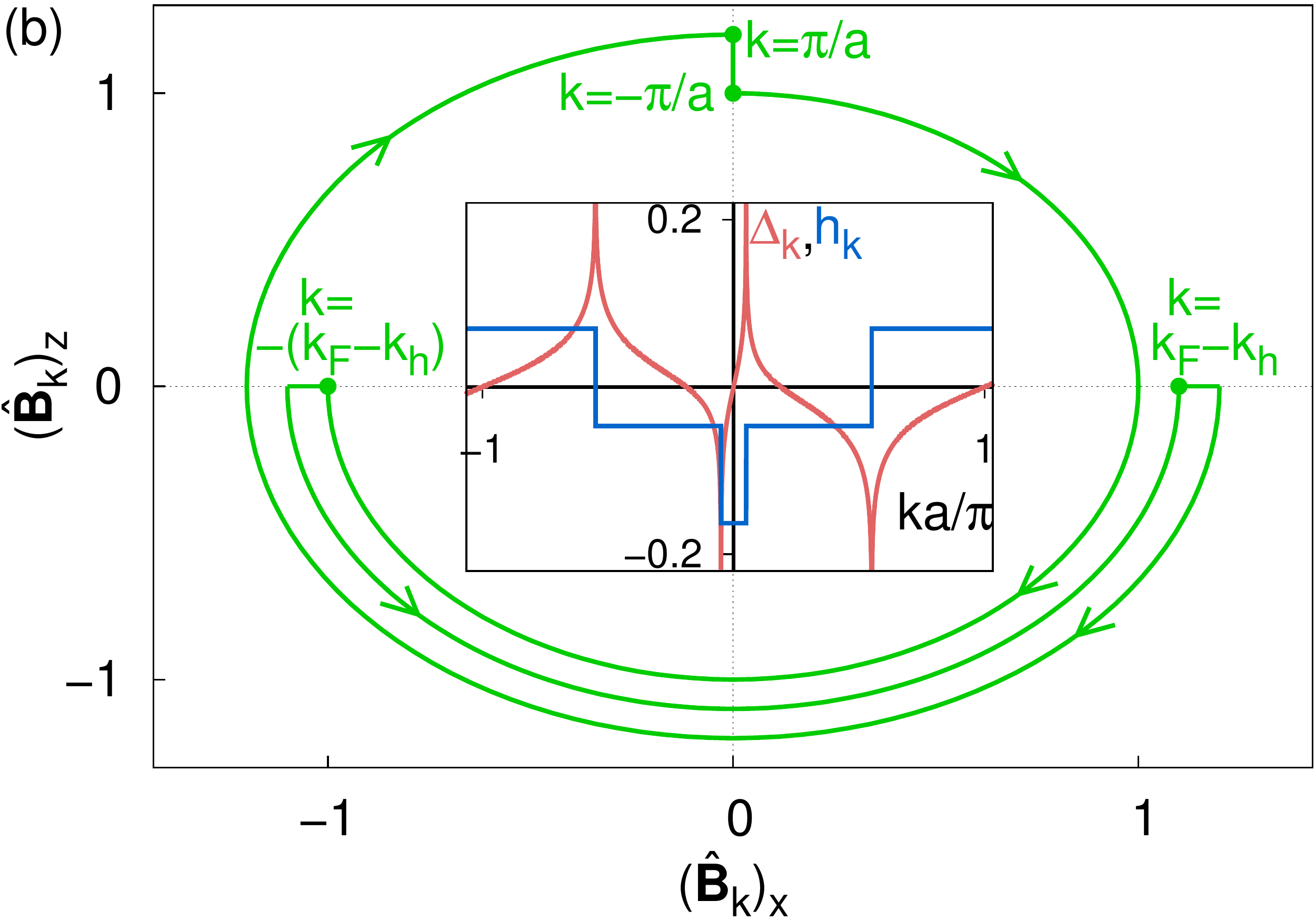}
\includegraphics[width=.32\textwidth]{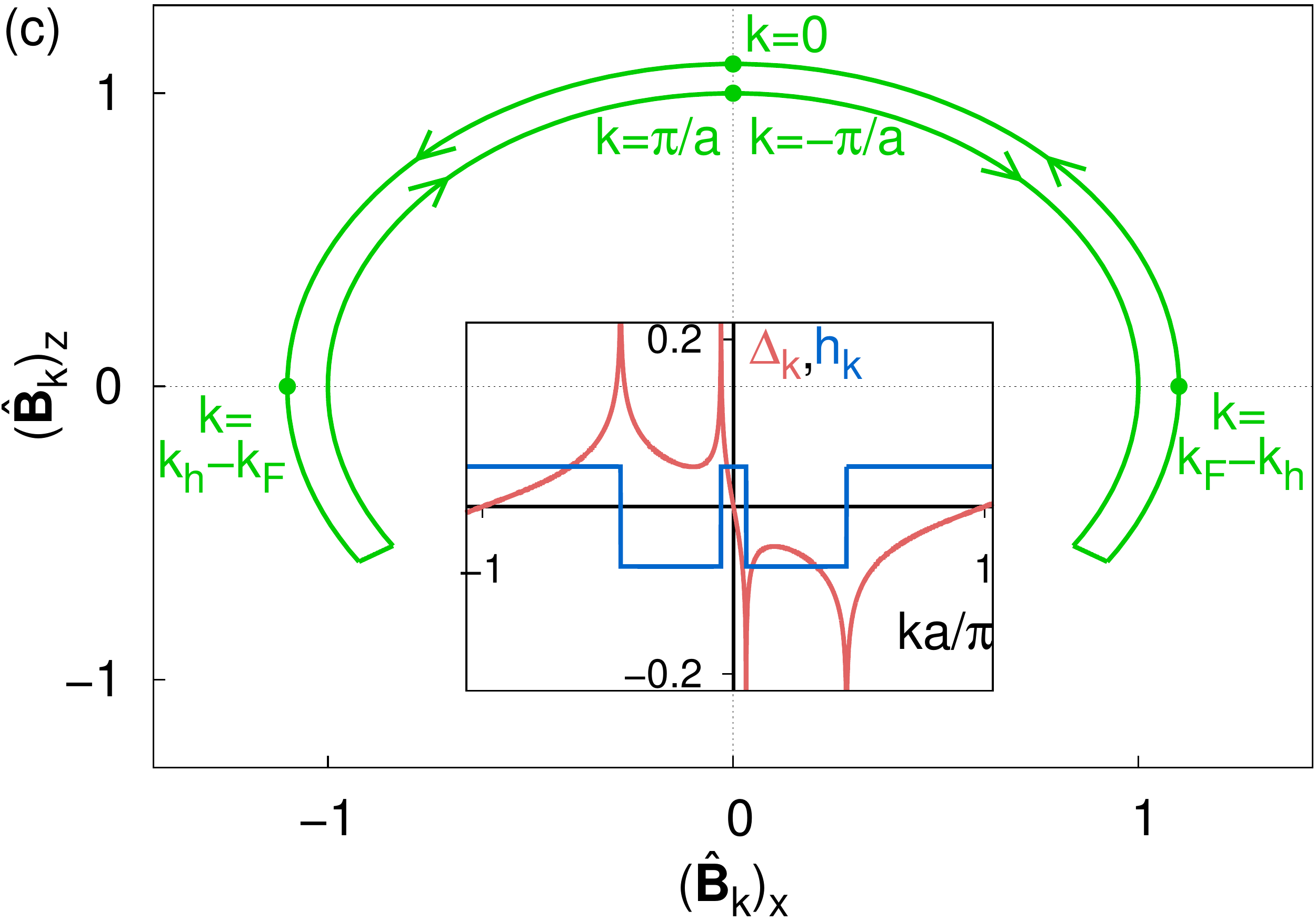}
\end{centering}
\caption{(a) Phase diagram for $k_h=0.1\pi/a$ and $\xi_0=\infty$ (black phase boundaries) or $\xi_0=15a$ (gray lines), with topological (T) and nontopological (N) phases. (Energies are given in units of $\Delta_0$.)  The yellow dashed line indicates the  Bragg point $k_F=k_h$ with exponentially localized Majorana states [$|\beta|<1$ in Eq.~(\ref{beta})]. For $\xi_0=\infty$, this coincides with the phase transition between the two- and single-channel phases. (b+c) Winding of the unit vector ${\bf\hat B}_k={\bf B}_k/B_k$ as $k$ is tuned across the Brillouin zone for (b) the single-channel and (c) the two-channel phase (partially shifted radially for visibility). While in the single-channel phase ${\bf\hat B}_k$ winds once around the origin; the winding is trivial in the two-channel phase, reflecting the topological phase transition. Insets: Dispersion $h_k$ and pairing $\Delta_k$ in the two phases. The two-channel dispersion has a second pair of Fermi points.
 }\label{fig:Beff}
\end{figure*}

{\em Model.---}We consider a linear chain of magnetic impurities which are located at positions $x_j = j a$ and form a planar spin helix ${\bf S}_j$, 
\begin{equation}
   (S_j)_x = S \cos 2k_hx_j \,\, , \,\, (S_j)_y = S \sin 2k_h x_j \,\, , \,\, (S_j)_z = 0. 
\end{equation}
In the limit that the Shiba states of the individual impurities are deep, i.e., that their energy $\epsilon_0$ is close to the center of the gap, the system can be effectively described by a tight-binding Bogoliubov--de Gennes (BdG) Hamiltonian ${\cal H} = h \tau_z + \Delta \tau_x$ \cite{pientka13}, where $\tau_i$ denotes Pauli matrices in particle-hole space, and $h$ as well as $\Delta$ are matrices in site space:
\begin{equation}
  h_{ij} = \epsilon_0 \delta_{ij} -  \Delta_0 (1 - \delta_{ij}) 
     \frac{\sin k_F r_{ij} }{k_F r_{ij}}e^{-r_{ij}/\xi_0}  \cos{k_hx_{ij}}  
     \label{heff_pi}
\end{equation}
and
\begin{equation}
  \Delta_{ij} = i \Delta_0 (1 - \delta_{ij}) 
       \frac{ \cos k_F r_{ij}}{k_F r_{ij}} e^{-r_{ij}/\xi_0}  \sin{k_h x_{ij}} . 
     \label{deltaeff__pi}
\end{equation} 
Here, $\Delta_0$ denotes the pairing strength in the host superconductor and $r_{ij} = |x_{ij}|$ with $x_{ij} = x_i - x_j$. Note that both, the hopping and pairing matrices are Hermitian, $h=h^\dagger$ and $\Delta = \Delta^\dagger$. Since the Shiba states are spin polarized, the pairing is effectively of $p$-wave nature, i.e., the pairing matrix is antisymmetric, $\Delta_{ij} = - \Delta_{ji}$. As explained above, this makes the Hamiltonian closely related to Kitaev's toy model \cite{kitaev2} except for the long-range nature of the hopping and pairing amplitudes. In view of the large ratio $\xi_0/a$, we consider the limit $\xi_0\to\infty$ in the following unless otherwise stated. 

This model has been discussed in detail in Ref.~\cite{pientka13} and we briefly review its topological properties before deriving the unusual localization properties of the Majorana states near the Bragg point. Figure \ref{fig:Beff}(a) reproduces the corresponding phase diagram as a function of Shiba bound state energy $\epsilon_0$ and Fermi wavevector $k_F$ \cite{pientka13}. The topological phase boundaries which appear in the phase diagram as diagonal lines occur when the chemical potential leaves the Shiba bands. These transitions are equivalent to those of the Kitaev chain and reflect a continuous gap closing and reopening. A different type of topological phase transition which emerges from the long-range coupling and has no analog in the Kitaev chain, occurs at the Bragg point $k_h=k_F$ (vertical lines) \cite{foot1}. This is a discontinuous transition associated with the (dis)appearance of an additional pair of Fermi points near $k=0$ [see insets of Figs.\ \ref{fig:Beff}(b) and (c)] modifying the system between effective single-channel and two-channel phases. In the nontopological two-channel phase, the cumulative hopping strength is finite across an even number of sites but vanishes across an odd number so that one can roughly think of the even and odd sites as two channels \cite{supp}. This happens for $k_F<k_h$, while even and odd sites are strongly coupled in the topological single-channel phase $k_F>k_h$. In some specific implementations, the RKKY interaction between the impurities is maximal at the wavevector $2k_F$, so that the helix wavevector realizes the Bragg point $k_h = k_F$ \cite{loss,braunecker09,simon08}. This would put the Shiba chain right at (for $\xi_0 \to \infty$) or near (for large but finite $\xi_0$) an unconventional topological critical point.

For a planar spin helix, the Shiba chain obeys chiral symmetry, $\{{\cal H},\tau_y\}=0$, which puts it in class BDI and in principle allows for a topological ${\mathbb Z}$-index \cite{Zindex}. To explore the discontinuous transition at the Bragg point more closely, we analyze the topological index of the two adjacent phases. To this end, we rewrite the Hamiltonian in momentum space, ${\cal H}_k= h_k \tau_z+\Delta_k\tau_x$, and determine the winding number of the two-component vector ${\bf B}_k=(\Delta_k,h_k)$ in the $xz$-plane as $k$ traverses the Brillouin zone from $-\pi/a$ to $\pi/a$ [see Figs.\ \ref{fig:Beff}(b) and \ref{fig:Beff}(c)]. This confirms the identification of the topologically trivial ($k_F<k_h$) and nontrivial ($k_F>k_h$) phases \cite{foot2}. 

\begin{figure*}[t]
\begin{centering}
\includegraphics[width=.32\textwidth]{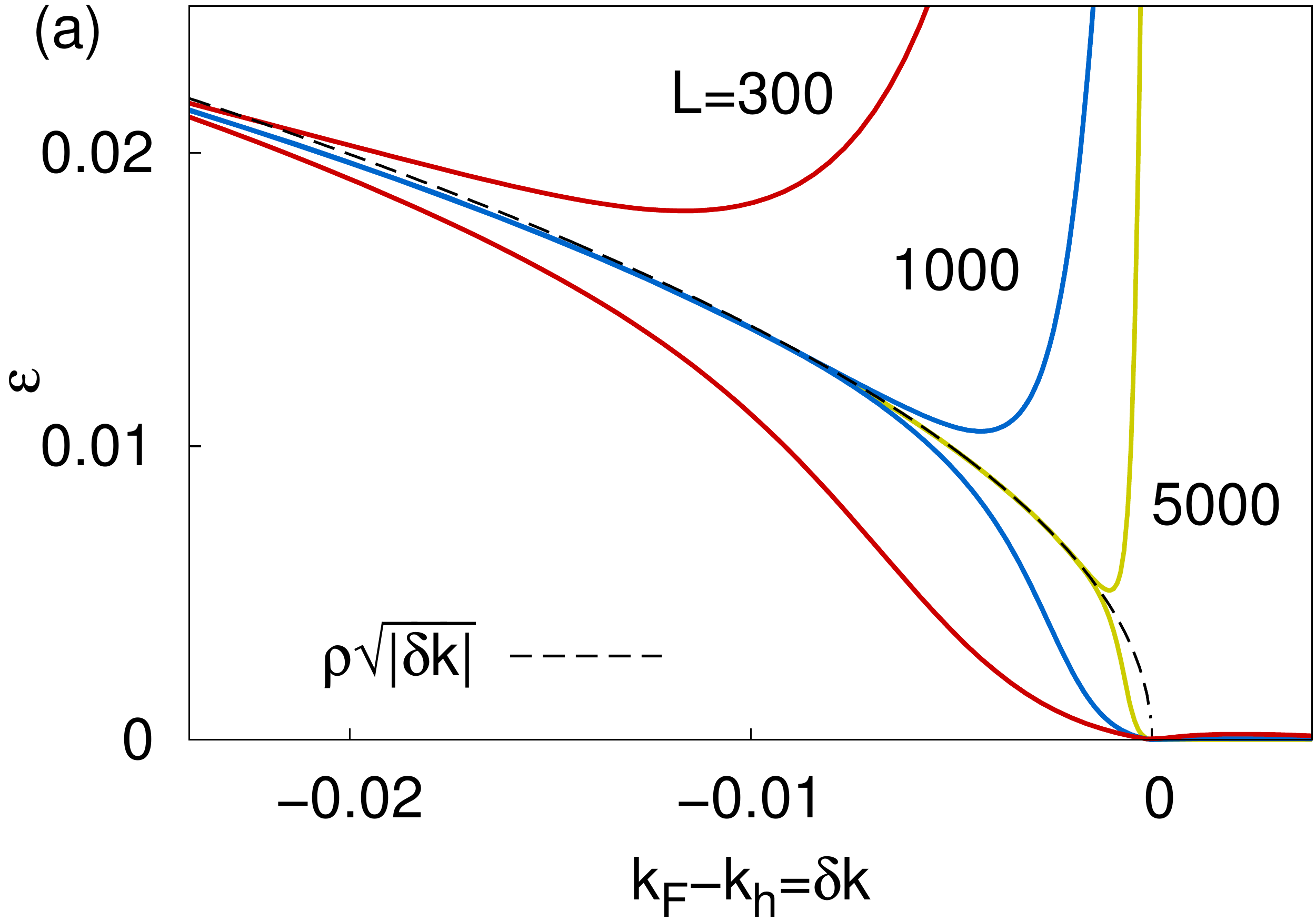}
\includegraphics[width=.32\textwidth]{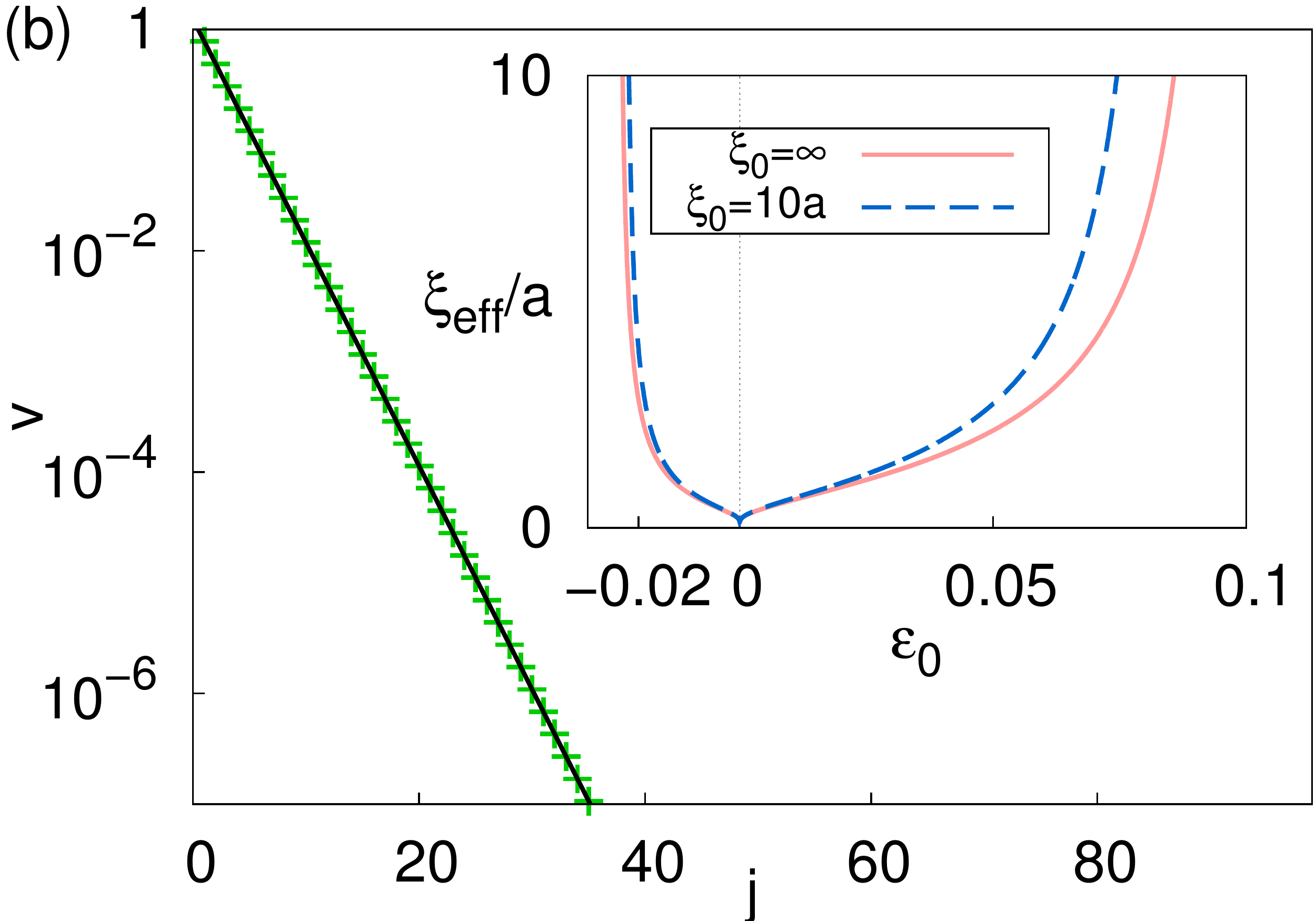}
\includegraphics[width=.32\textwidth]{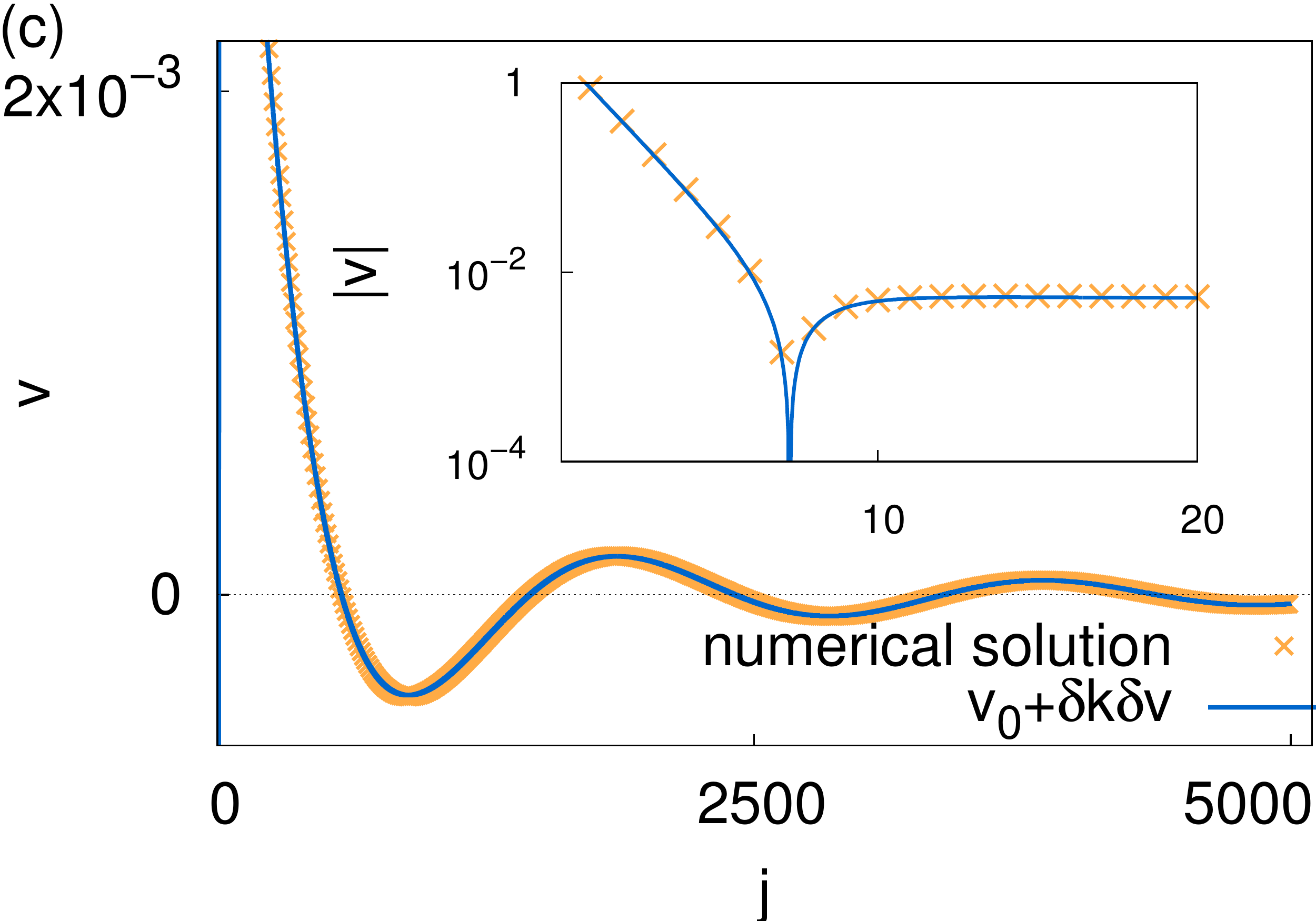}
\end{centering}
\caption{(a) Energy of the two positive-energy subgap states (in units of $\Delta_0$) in the nontopological two-channel phase ($\delta k<0$) near the Bragg point, for $\xi_0=\infty$ and various chain lengths $L$. As $L\to \infty$, the two states become degenerate with energy $\sim \sqrt{|\delta k|}$ near the phase transition. At the critical point, one subgap state merges discontinuously with the quasiparticle continuum. For finite $L$, the discontinuity is smeared and the degeneracy is lifted on the scale $1/L$. (b) Majorana wavefunction $v_j$ at the Bragg point $k_F=k_h$. The exact numerical solution of the BdG Hamiltonian (green crosses) agrees with the analytical solution (black line) in Eq.~(\ref{beta}). Inset:  Localization length $\xi_{\rm eff}=a/\ln |\beta^{-1}|$ along the yellow line in the phase diagram in Fig.~\ref{fig:Beff}(a). The localization length is of order $a$ and decreases with increasing coherence length $\xi_0$. (c) Majorana wavefunction $v_j$ for $k_F=k_h + \delta k$ with $\delta k=0.003/a$. The numerical solution of the BdG Hamiltonian (orange crosses) agrees with the analytical solution (blue line) as obtained by numerical evaluation of the inverse Laplace transform in Eq.\ (\ref{inverse_laplace}). Inset: Blowup near the end of the chain emphasizing the initial exponential decay. Parameters: $\epsilon_0=0.03\Delta_0$, $k_h=0.1\pi/a$, and $k_F=4.1\pi/a+\delta k$.
}\label{fig:wv}
\end{figure*}

The transition between these phases at the Bragg point is reflected in the subgap states of long but finite chains. Their energies near $k_F=k_h$ are shown in Fig.~\ref{fig:wv}(a). In the two-channel phase, one finds two subgap states for each end. These can be thought of as the hybridized Majorana states of the two channels. As $k_F\to k_h$, one subgap state merges with the quasiparticle continuum due to coupling with the opposite end of the chain, while the other approaches zero energy and connects smoothly with the Majorana end state in the topological phase. Thus, right at the Bragg point, there is exactly one zero-energy state for each end of the chain. We now turn to an analytical theory of the Majorana bound state and the subgap spectrum both at and near the Bragg point.

{\em Majorana bound state at the Bragg point $k_F=k_h$.---}We exploit the chiral symmetry of the Hamiltonian and rotate it into the Majorana basis in which ${\cal H}$ becomes purely off-diagonal in particle-hole space \cite{Rotation}. This is effected by a $\pi/2$-rotation about the $x$-axis which transforms $\tau_z\to -\tau_y$ and keeps $\tau_x$ unchanged so that ${\cal H} = -h\tau_y + \Delta \tau_x$. Now, the equations for the zero-energy Majorana states with BdG spinor $(u,v)$ take the simple form 
\begin{equation}
     u=0 \,\,\,\,\,\,\,\, \hbox{\rm and} \,\,\,\,\,\,\,\,  (ih + \Delta)v=0 
\label{leftmajo}
\end{equation}
for the Majorana localized at the left end of the chain, and $v=0$ and  $(-ih + \Delta)u=0$ for the Majorana localized at the right end. (Note that this consistently neglects finite-size effects). Specifying to the left-end states for definiteness, one readily finds for ${\cal H}_{12}= ih + \Delta$, from Eqs.\ (\ref{heff_pi}) and (\ref{deltaeff__pi}), that $({\cal H}_{12})_{jj} = i \epsilon_0$ and 
\begin{equation}
    ({\cal H}_{12})_{ij} = - \frac{ i \Delta }{k_F r_{ij}}e^{-r_{ij}/\xi_0}  \sin (k_F r_{ij} - k_h x_{ij})\label{H12}
\end{equation}
for $i\neq j$. Since $r_{ij}=|x_{ij}|$, the Bragg point $k_h = k_F$ has the remarkable property that ${\cal H}_{12}$ is a triangular matrix. This property immediately allows us to solve Eq.\ (\ref{leftmajo}) by the ansatz $v_j = \beta^j$. Here, $j$ enumerates the sites starting with the left end of the chain. Indeed, with this ansatz, all components of the equation ${\cal H}_{12} v = 0$ reduce to the same condition for $\beta$. Solving this condition, we find
\begin{equation}
   \beta  = \frac{e^{a/\xi_0}\sin(k_F a\epsilon_0/\Delta)}{\sin[2k_F a + k_F a\epsilon_0/\Delta]}.\label{beta}
\end{equation}
Obviously, this provides an exponentially localized Majorana solution as long as $|\beta|<1$. One can convince oneself that this condition is satisfied wherever the line $k_F=k_h$ is inside the topological phase for finite $\xi_0$. This region is marked by a yellow dashed line in Fig.~\ref{fig:Beff}(a). As shown in Fig.\ \ref{fig:wv}(b), this exact analytical result is in excellent agreement with numerical simulations.

This constitutes the central result of this work with remarkable implications: 
(i) Helical Shiba chains display an unconventional topological critical point at or in the immediate vicinity of the Bragg point $k_F=k_h$. (ii)
At the Bragg point, they have Majonana end states which are exponentially localized even though the Hamiltonian allows for long-range
hopping and pairing along the chain. (iii) The localization length $\xi_{\rm eff}=a/\ln |\beta^{-1}|$ of the Majorana states is set by the spacing $a$ between the magnetic impurities and thus much shorter than the coherence length $\xi_0$ of the superconducting host. (iv) The Majorana end states at the Bragg point remain well-defined and exponentially localized even in the limit $\xi_0\to\infty$ where the Bragg point coincides with the topological critical point. (v) We will see below that away from the Bragg point, the Majorana wavefunctions develop a power-law tail in addition to the initial exponential decay. 

Physically, the strong localization for $k_h=k_F$ can be traced back to Bragg reflection. Similar to a Bragg mirror, the resonance between the oscillations of the Shiba states and the spin helix leads to destructive interference in one direction which neutralizes the long-range coupling. This explains that the localization length becomes of the order of the lattice spacing. More explicitly, the hopping and pairing terms are generically of the same order, but their relative magnitude depends sensitively on the wavevectors $k_F$ and $k_h$. At the resonance $k_F=k_h$, hopping and pairing between two arbitrary sites have equal magnitude but differ in parity. Hopping to left and right has the same sign and is thus even, whereas pairing is odd with opposite signs for the two directions. 

{\em Topological phase.---}When tuning away from the Bragg point, Bragg reflection is no longer perfect and the long-range character of the model is partially recovered. As a result, the wavefunction acquires a tail with a slow power-law decay as we will now show for the immediate vicinity of the Bragg point (in agreement with earlier numerical results \cite{pientka13}). Here we first focus on the topological phase ($k_F = k_h + \delta k$ with $\delta k$ small and positive) and return to the nontopological phase ($k_F = k_h + \delta k$ with $\delta k$ small and negative) further below. 

For $\delta k$ small and positive, the matrices ${\cal H}_{12}$ and ${\cal H}_{21}$ are no longer triangular, but we still expect a localized Majorana state in a semi-infinite chain. We expand the eigenvalue problem to linear order in $\delta k$ and show that it reduces to an integral equation in a suitably taken limit when setting $\xi_0 \to \infty$. The integral equation can then be solved by standard methods. To first order in $\delta k$, we rewrite ${\cal H}_{12}v=0$ as
\begin{align}
 (M+\delta M)(v_0+\delta k \delta v)=0 \label{eigenproblem}
\end{align}
(see \cite{supp} for numerical support of this expansion). Here, $M$ is the upper triangular part of ${\cal H}_{12}$ with $M_{ii}=i\epsilon_0$, $M_{ij}=(-i\Delta/k_F) \sin( Kr_{ij})/r_{ij}$ for $i<j$ in terms of $K=k_F+k_h$, and $\delta M$ is the lower triangular part with $\delta M_{ij}=(-i\Delta/k_F) \sin(\delta k r_{ij})/r_{ij}$ for $i>j$. To zeroth order in $\delta k$, we obtain $Mv_0=0$ and thus $v_0$ coincides with the exponentially decaying solution at the Bragg point.

Next, we rewrite $\delta M$ as $\delta M_{i+j,i}=(-i\Delta/k_F)\delta k\sin y_j/y_j$ with $y_j=\delta k aj$. Thus $\delta M_{ij}$ varies only on large scales $r_{ij}\sim 1/\delta k$ and we can take a continuum limit by considering $\delta k\to 0$ while keeping $y_j$ fixed. In this limit, $\delta M$ converges to a continuous matrix as a function of $y_j\to y$ and correspondingly, $\delta v$ should also have a well-defined continuum limit $\delta v\to \delta v(y)$ as a function of the scaled variable $y_j$. The existence of this continuum limit is readily confirmed by numerics and to linear order in $\delta k$, Eq.~(\ref{eigenproblem}) yields an integral equation for $\delta v(y)$ \cite{supp},
\begin{eqnarray}
 A \delta v(y) +  \int_0^y dz \frac{\sin(y-z)}{y-z}\delta v(z) = -B\frac{\sin y}{y}.
\label{IntEq}
\end{eqnarray}
Here we defined $A =  F(k_F+k_h)-\frac{k_Fa\epsilon_0}{\Delta}$, $B=a\, {\rm sgn}\, \beta [({1+\beta})/({1-\beta})]^{1/2}$, and  $F(x) =\arctan\cot(x/2)$. This integral equation can be solved in a standard manner by Laplace transform ${\cal L}$ which yields \cite{supp} 
\begin{eqnarray}
  \delta v(y) =-{\cal L}^{-1}\left[ \frac{B\,{\rm arccot}\, s}{A+{\rm arccot}\, s}\right] 
 \underset{y\to\infty}{\sim} -4 AB \frac{\sin y}{y\ln^2 y} .
\label{inverse_laplace}
\end{eqnarray}
Corrections to the asymptote are suppressed by factors of $1/\ln y$. Although our analytical analysis focuses on the vicinity of the Bragg point, the asymptotic decay is characteristic of the Majorana states in the entire topological phase when $\xi_0=\infty$ as previously established numerically \cite{pientka13}. Figure \ref{fig:wv}(c) shows that the analytical solution (\ref{inverse_laplace}) is in excellent agreement with numerical results.

{\em Nontopological phase.---}When the Bragg point $k_F=k_h$ falls into the topological phase for finite $\xi_0$ [i.e., along the yellow line in Fig.~\ref{fig:Beff}(a)], the nontopological side of the phase transition ($\delta k<0$) can be understood as an effective two-channel wire \cite{pientka13}. Thus, this phase exhibits two subgap states for each end. In long chains, $L\to \infty$, their wavefunction and energy can be obtained analytically by an extension of the technique used for $\delta k>0$. Here, we sketch the results and defer details to the Supplementary Material \cite{supp}. 

The two positive-energy subgap states become degenerate for large $L$ and to leading order in $|\delta k|$, we find that their energy scales as $\epsilon\sim |\delta k|^{1/2}$, consistent with Fig.~\ref{fig:wv}(a). Similarly, the Nambu wavefunction $(u,v)$ for the state at the left end \cite{foot3} has $u_i\sim |\delta k|^{1/2}\exp(-\cot A |\delta k|ai)$ and $v\sim v_0+|\delta k|\delta v$, with $v_0$ the exponential solution at the critical point and $\delta v$ a power-law tail $\sim 1/y\ln^2 y$ as for $\delta k >0$. The electron component $u$ decays exponentially with a decay length which diverges for $\delta k=0$, reflecting the phase transition and the disappearance of one subgap state. The hole component $v$ smoothly evolves into the Majorana bound state on the topological side. 

{\em Conclusions.---}As a consequence of the long-range coupling, helical Shiba chains display an unconventional topological critical point at the Bragg point $k_F=k_h$. We show that for finite chains, the Majorana end states persist at the critical point and display remarkable localization properties. By the competition between Bragg reflection and long-range coupling, the Majorana end states are exponentially localized at the critical point but develop a power-law tail in the topological phase. This contrasts in an interesting way with the decay of correlations around conventional critical points.

This is also an encouraging prediction for experiment, as the exponential localization at the critical point is on the scale of the lattice spacing and entirely unrelated to a topological gap. Thus, the Majorana end states may remain well localized even in chains whose length is comparable to the coherence length of the host superconductor. At the same time, the power-law localization within the topological phase raises interesting questions with regard to its stability against perturbations such as disorder.  

{\em Acknowledgments.---} We thank Ali Yazdani for stimulating discussions and Roman Lutchyn and Jelena Klinovaja for useful comments. We acknowledge financial support by the Helmholtz Virtual Institute ``New states of matter and their excitations," SPP1285 of the Deutsche Forschungsgemeinschaft, the Studienstiftung d.\ dt.\ Volkes, and NSF DMR Grants 0906498 and 1206612. We are grateful to the Aspen Center for Physics, supported by NSF Grant No.\ PHYS-106629, for hospitality while this line of work was initiated.

\onecolumngrid
\clearpage
\section*{Supplementary Material for\\ Unconventional topological phase transitions in helical Shiba chains }

\twocolumngrid

\subsection*{Origin of the two-channel phase}
\begin{figure}[t]
\begin{centering}
\includegraphics[width=.46\textwidth]{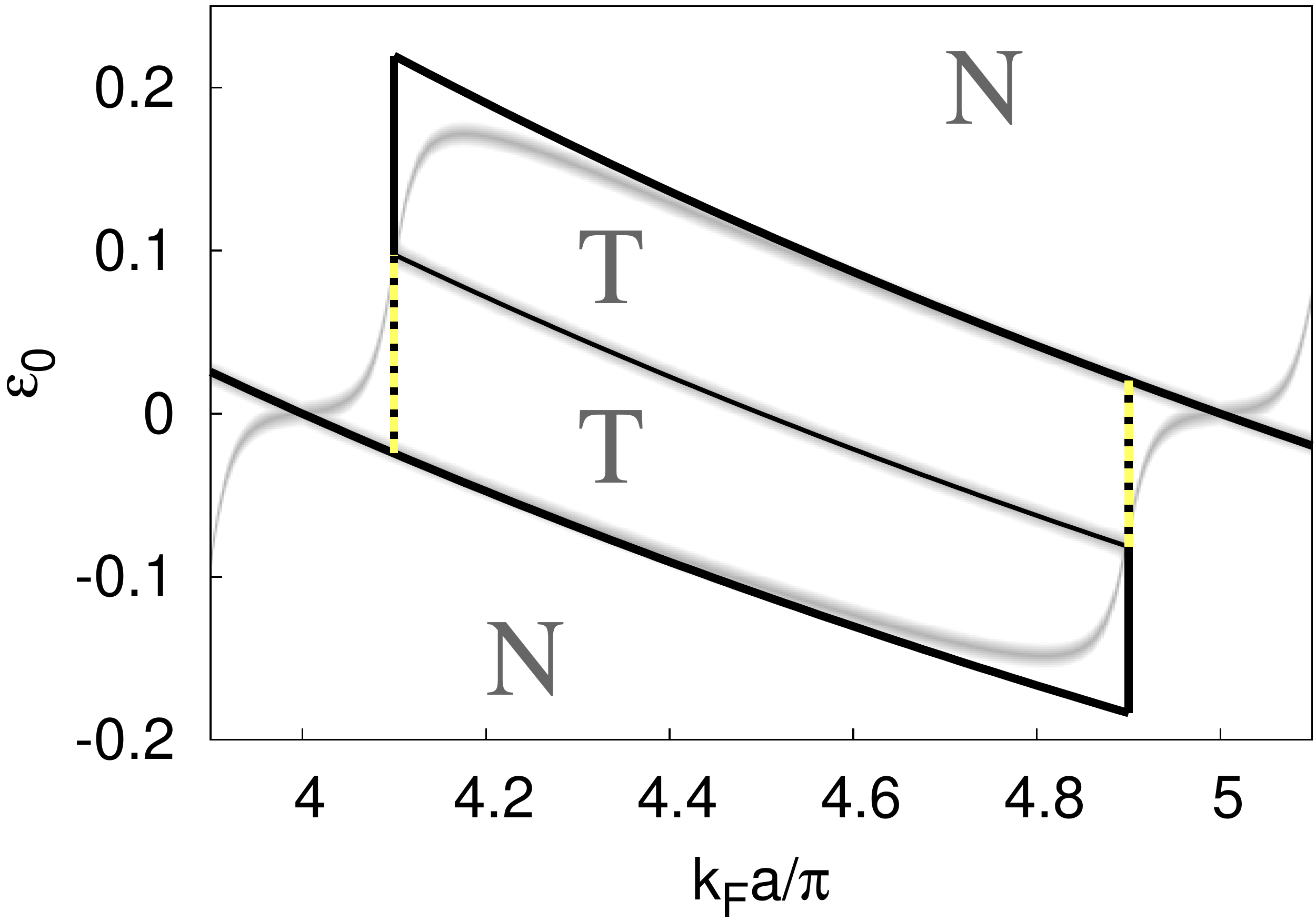}
\includegraphics[width=.46\textwidth]{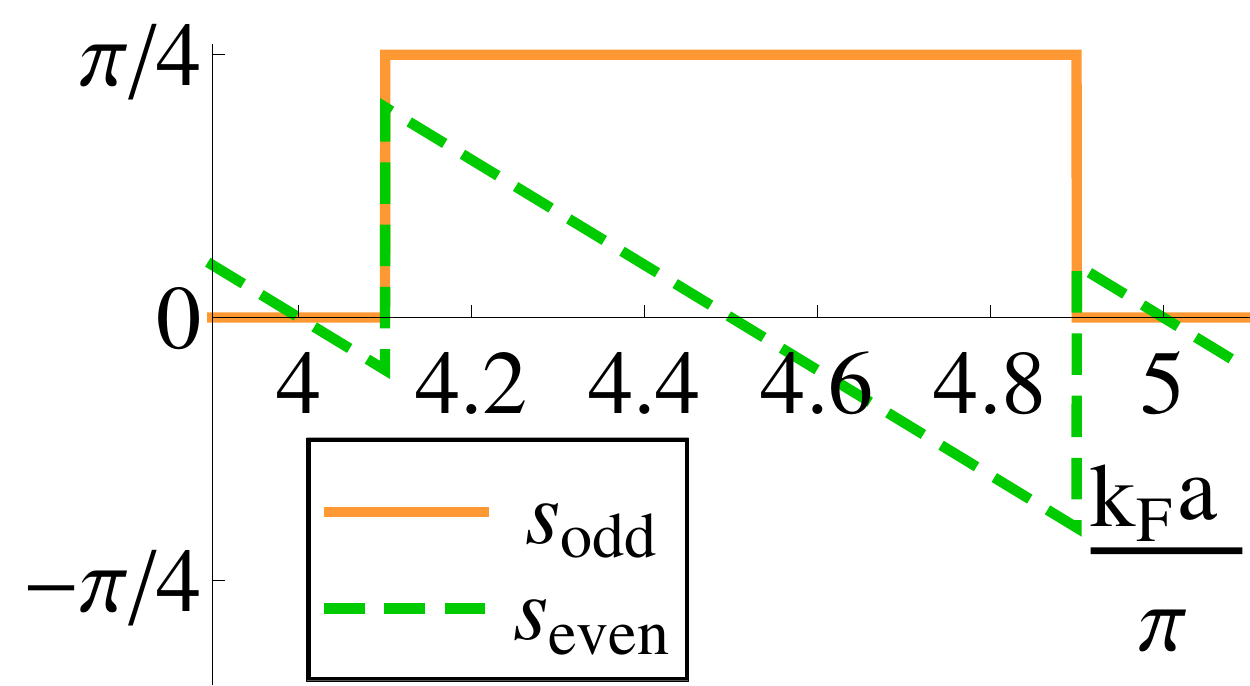}
\end{centering}
\caption{Phase diagram from Fig.~\ref{fig:Beff}(a) of the main text and corresponding cumulative hopping amplitudes across an odd and even numerber of lattice sites ($\xi_0=\infty$, $k_h=0.1\pi/a$): $s_{\rm odd}=\sum_{n}\sin[ k_Fa (2n+1)]\cos [k_ha(2n+1)]/(2n+1)$ and $s_{\rm even}=\sum_{n}\sin (2k_F an)\cos( 2k_ha n)/2n$. While $s_{\rm even}$ is always nonzero except for special points, $s_{\rm odd}$ changes from zero to $\pi/4$ at the transition between two-channel ($k_F<k_h$) and single-channel phase ($k_F>k_h$).
}\label{fig:sum}
\end{figure}

Roughly speaking, the two-channel phases arises because the even and odd sites of the chain form two independent channels. This can be traced back to the fact that for some ranges of the wavevectors $k_F$ and $k_h$, hopping across an even number of lattice sites becomes stronger than across an odd number. To make this intuitive understanding more plausible, we consider the lattice Fourier transform $h_{k=0}=\sum_n h_{i,i+n}$ for the wavevector $k=0$, at which the dispersion changes abruptly when $k_F=k_h$. The contribution to $h_{k=0}$ from hopping an odd number of sites is $\sum_{n}\sin k_Fa (2n+1)\cos k_ha(2n+1)/(2n+1)$. This vanishes in the two-channel phase $k_F<k_h$ and equals $\pi /4$ in the single-channel phase $k_F>k_h$. The sum over all even hopping terms, on the other hand, varies between $-\pi/4$ and $\pi/4$. The cumulative hopping amplitudes for even and odd sites are illustrated in Fig.~\ref{fig:sum}. Thus, the phase transition occurs because hopping between even and odd sites becomes ineffective for $k_F<k_h$ and the chain effectively decomposes into two channels. This two-channel phase arises as a consequence of the long-range nature of the model. For finite $\xi_0$, the long-range hopping is cut off by the coherence length and the two-channel phase is restricted to a smaller region of parameter space.

\begin{figure}[t]
\begin{centering}
\includegraphics[width=.46\textwidth]{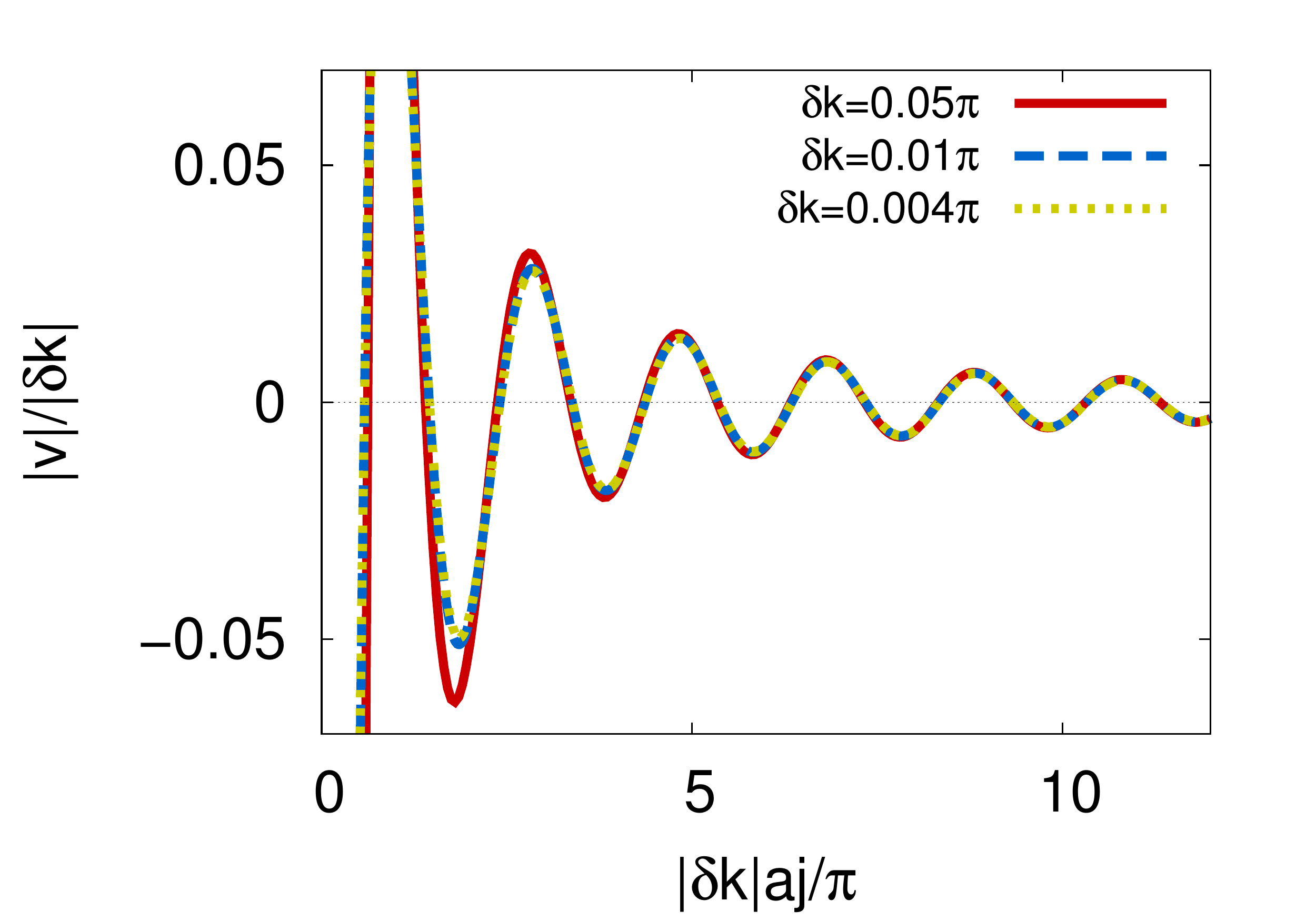}
\end{centering}
\caption{Scaling behavior of the Majorana wavefunction for $\delta k >0$. In the limit $\delta k\to0$ the slowly decaying tail $\delta k\delta v$ only depends on the scaled position variable $y_j=\delta ka j$. Due to the scaled axes all curves collapse. Note that the exponentially decaying zeroth-order term $v_0$ only dominates on the first few sites, which cannot be resolved in this graph. Similar results are obtained for $\delta k<0$. The parameters used for numerical calculations in the Supplementary Material are all chosen as in Fig.~\ref{fig:wv} of the main text.
}\label{fig:wv-scale}
\end{figure} 

\subsection*{Exponentially localized Majorana bound state at the Bragg point}

We now provide further details of the derivation of Eq.~(\ref{beta}) of the main text. At the Bragg point $k_F=k_h$, the off-diagonal block of the rotated Hamiltonian given in Eq.~(\ref{H12}) reads
\begin{align}
 ({\cal H}_{12})_{ij}=
 \begin{cases}
 -\frac{i\Delta}{k_Fa}e^{-(j-i)a/\xi_0}\frac{\sin [2k_Fa(j-i)]}{j-i}& j>i,\\
  i\epsilon_0& j=i,\\
 0&  j<i.
 \end{cases}
\end{align}
Thus ${\cal H}_{12}$ is triangular and constant along the diagonals. In order to find Majorana end states we seek a solution of ${\cal H}_{12}v=0$ which gives the set of equations
\begin{align}
0&= \sum_{j=1}^\infty({\cal H}_{12})_{ij}v_j\\
 &=i\epsilon_0 v_i-\frac{i\Delta}{k_Fa}\sum_{j=i}^\infty e^{-(j-i)a/\xi_0}\frac{\sin 2k_Fa(j-i)}{j-i}v_j.
\end{align}
Using the ansatz $v_j=\beta^j$ all of these equations reduce to the same condition
\begin{align}
 i\epsilon_0 =\frac{i\Delta}{k_Fa}\sum_{j=1}^\infty e^{-ja/\xi_0}\frac{\sin (2k_Faj)}{j}\beta^j.
\end{align}
We can rewrite this as 
\begin{align}
 i\epsilon_0 =\frac{\Delta}{2k_Fa}\sum_{j=1}^\infty\left[ \frac{(\beta e^{-a/\xi_0+i2k_Fa})^j}{j}-{\rm c.c.}\right]
\end{align}
and using the identity $\sum_{j=1}^\infty x^j/j=-\ln(1-x)$ we obtain
\begin{align}
 i\epsilon_0 &=\frac{\Delta}{2k_Fa}\ln\left[\frac{1-\beta e^{-a/\xi_0+i2k_Fa}}{1-\beta e^{-a/\xi_0-i2k_Fa}}\right]\\
 &=\frac{i\Delta}{k_Fa}\arctan\frac{\beta e^{-a/\xi_0}\sin(2k_Fa)}{1-\beta e^{-a/\xi_0}\cos(2k_Fa)}.
\end{align}
Rearranging this as 
\begin{align}
\beta &e^{-a/\xi_0}[ \sin(2k_Fa)\cos \frac{k_Fa\epsilon_0}{\Delta}+\cos(2k_Fa)\sin \frac{k_Fa\epsilon_0}{\Delta} ]\nonumber\\
&= \sin \frac{k_Fa\epsilon_0}{\Delta},
\end{align}
and solving for $\beta$ finally yields Eq.~(\ref{beta}).

\subsection*{Analytical solution near the Bragg point for $\xi_0\to \infty$}

Here we derive the corrections to the exponential Majorana wavefunction in the limit $\xi_0\to \infty$ when tuning slightly away from the phase transition at $\delta k=0$. To this end we expand the eigenproblem in $|\delta k|$ and construct a continuum limit in which the matrix equations for the wavefunction transform into integral equations that can be solved by standard methods. We present the solution for the more involved case $\delta k<0$ (nontopological side). The calculation for the topological side $\delta k>0$ is contained in the derivation presented here and can be obtained by simply setting $u$ and $\epsilon$ in Eq.~(\ref{eigenprob_ndk}) to zero. On the nontopological side, there are two degenerate subgap states which exist at a finite energy. The eigenproblem reads
\begin{align}
 \begin{pmatrix}
  0&M+\delta M\\
  M^\dag+\delta M^\dag &0
 \end{pmatrix}\binom{u}{v}=\epsilon\binom{u}{v}\label{eigenprob_ndk}
\end{align}
with
\begin{align}
 M_{i,j}&=
 \begin{cases}
 \frac{-i\Delta}{k_Fa}\frac{\sin[ Ka (j-i)]}{j-i}&\ j>i\\
 i\epsilon_0& \ j=i\\
  0& \ j<i
 \end{cases}\\
 \delta M_{i,j}&=
 \begin{cases}
 0&\ j\geq i\\
  \frac{-i\Delta}{k_Fa}\frac{\sin [\delta k a(i-j)]}{i-j}& \ j<i
 \end{cases}.
\end{align}
and $K=2k_F+\delta k$. The two degenerate subgap states are localized at opposite ends of the chain by an appropriate choice of basis. For definiteness we consider the Nambu spinor $(u,v)$ to be localized at the left end and solve the eigenproblem for a semi-infinite chain. We start by expanding $v$ to first order in $|\delta k|$,
\begin{align}
 v=v_0+|\delta k|\delta v.
\end{align}
As detailed in the main text we find that in the limit $\delta k\to 0$ with $y_j={\rm const}$, the correction $\delta v$ depends only on the scaled position variable $y_j=|\delta k|aj$. In this limit, the position variable becomes continuous $y_j\to y$ and the matrix equations (\ref{eigenprob_ndk}) transforms into integral equations. The expansion and the scaling behavior of $\delta v$ is corroborated by numerical results in Fig.~\ref{fig:wv-scale}. 

We can now obtain the scaling of the energy $\epsilon$ with $\delta k$ by eliminating $u$ from Eq.\ (\ref{eigenprob_ndk}),
\begin{align}
(M^\dag+\delta M^\dag)(M+\delta M)v=\epsilon^2 v. \label{eigprob_square}
\end{align}
As $\epsilon\to 0$ with $\delta k\to 0^-$ (which is supported by numerics in Fig.~\ref{fig:wv}(a) of the main text), we obtain $Mv_0=0$ in zeroth order, which is the eigenproblem at the Bragg point, i.e.,
\begin{align}
v_{0,j}=\mathcal{N}_1\beta^j,
\end{align}
with a normalization constant $\mathcal{N}_1$ and $\beta$ given by Eq.~(\ref{beta}). Normalization readily yields ${\cal N}_1=(1-\beta^2)^{1/2}/|\beta|$ for $\delta k>0$. When $\delta k<0$, the hole component $u$ also enters into the normalization condition and we will determine ${\cal N}_1$ later. 

\begin{figure}[tb]
\centering
\includegraphics[width=.45\textwidth]{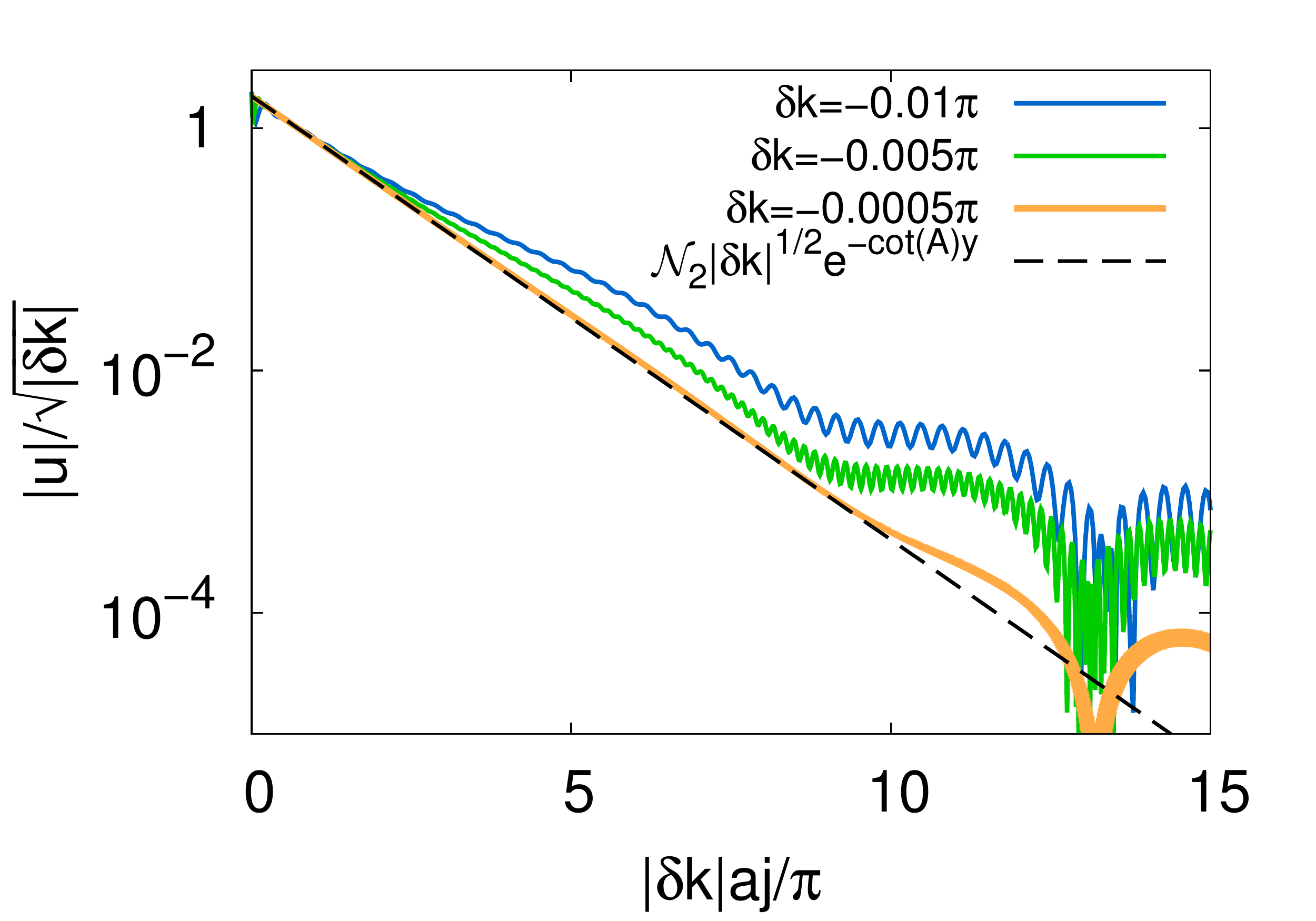}
\caption{Nambu component $u$ for various $\delta k<0$. In the limit $\delta k \to 0$ the dominating contribution near the end of the chain is an exponential function which is well-described by our analytical solution (dashed line) given in Eq.~(\ref{usol}). The results also show the scaling $|\delta k|^{1/2}u_0(|\delta k|aj)$. Further inside the chain $u$ has a slowly decaying tail of order $|\delta k|^{3/2}$, which can be described by a similar power-law decay as $\delta v$. Since the tail of $u$ is higher-order term we do not consider it further. 
}
\label{fig:wv-scale-u-exp}
\end{figure}

From the next order in Eq.~(\ref{eigprob_square}), we find that the energy scales as $\epsilon=\rho \sqrt{|\delta k|}$ with some constant $\rho$. This result is consistent with the numerical results in Fig.~\ref{fig:wv}(a) of the main text. Since $v_0$ corresponds to the single solution at the Bragg point, $u$ cannot have a zero-order term. In fact, due the $\sqrt{|\delta k|}$-dependence of $\epsilon$, we have to expand $u=|\delta k|^{1/2}u_0+O(|\delta k|^{3/2})$. Also $u_0$ is a function of $y_i$ if we ignore terms at the very end of the chain where $v_0$ is dominating. The scaling $|\delta k|^{1/2}u(|\delta k|ai)$ is confirmed by numerics as shown in Fig.~\ref{fig:wv-scale-u-exp}.

The next order $\sim |\delta k|^{1/2}$ of Eq.~(\ref{eigenprob_ndk}) is given by
\begin{align}
|\delta k|^{1/2}\left( M^\dag u_0+\delta M^\dag u_0-\rho v_0\right)=0.
\end{align}
As mentioned above we ignore boundary terms of $u_0$ since they are of subleading order in $|\delta k|$. Hence we omit the last term $\sim v_0$ in the previous equation and take the limit $|\delta k|\to 0$ with $y_i={\rm const.}$ of the remaining terms by introducing the continuous position variable $y_i\to y$. This yields 
\begin{align}
[(M^\dag+\delta M^\dag ) u_0]_i = -\frac{i\Delta}{k_Fa}\sum_{j=1}^{i-1} \frac{\sin [Ka(i-j)]}{i-j}u_{0}(y_j)\nonumber \\
+i\epsilon_0 u_{0}(y_i) -\frac{i\Delta}{k_Fa}\sum_{j=i+1}^{\infty} \frac{\sin [\delta ka(j-i)]}{j-i}u_{0}(y_j)=0.\label{deltak_1/2}
\end{align}
The first sum can be rewritten as
\begin{align}
\sum_{j=1}^{i-1} \frac{\sin (Kaj)}{j}u_{0}(y_i-y_j). 
\end{align}
When $|\delta k|$ is small, $u_0(y_i-y_j)$ becomes a slowly varying function and the sum is dominated by the decay $\sin (Kj)/j$, i.e., only terms with $j\lesssim O(1/K)$ are relevant. Thus in the limit $\delta k\to 0$, $i\to\infty$ with $y_i={\rm const.}$ we can set $y_j\to 0$ and obtain
\begin{align}
  \sum_{j=1}^{i-1} \frac{\sin [Ka(i-j)]}{i-j}u_{0}(y_j) &\to u_0(y)\sum_{j=1}^\infty\frac{\sin (K j)}{j}\\
  &= u_0(y)\arctan\cot(K/2).
\end{align}
The other terms have a straightforward continuum limit and we obtain
\begin{align}
Au_0(y) +{\rm sgn}(\delta k )\int_{y}^{\infty} dz \frac{\sin(z-y)}{z-y}u_0(z)=0,
\end{align}
where $A=\arctan\cot(K/2)-k_Fa\epsilon_0/\Delta$ as in the main text. The solution of this homogeneous integral equation for $\delta k<0$ is given by \cite{polyanin}
\begin{align}
  u_0(y)=\mathcal{N}_2e^{-\cot (A) y },\label{usol}
\end{align}
where the normalization factor ${\cal N}_2$ will be determined below. This result agrees well with a numerical solution of the tight-binding BdG Hamiltonian in the limit $\delta k\to 0$, as shown in Fig.~\ref{fig:wv-scale-u-exp}.

\begin{figure}[tb]
\centering
\includegraphics[width=.23\textwidth]{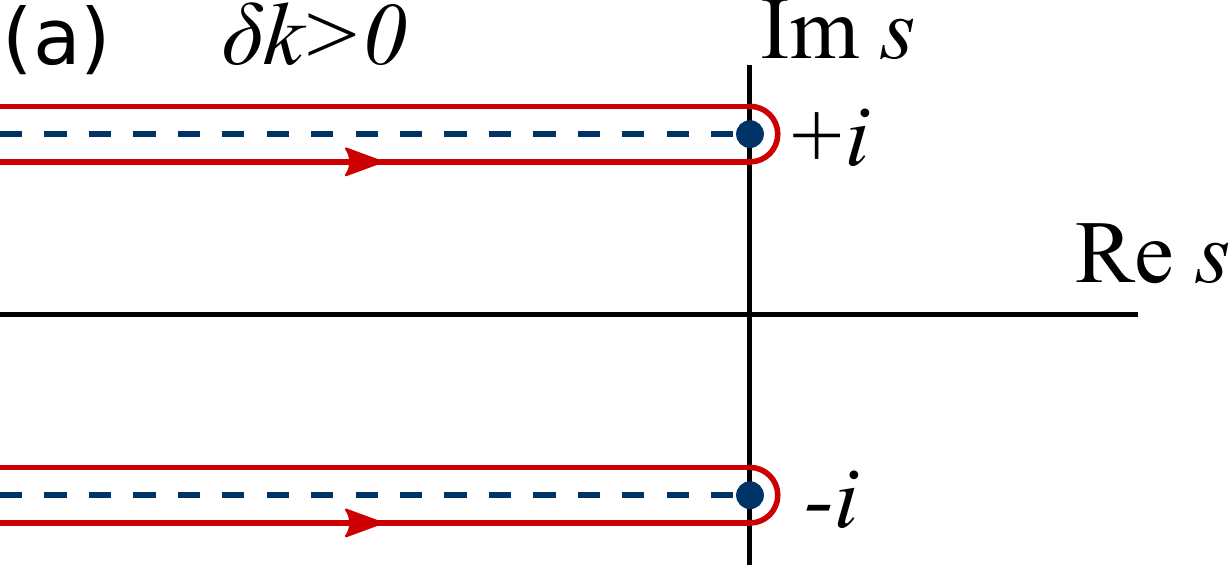}
\includegraphics[width=.23\textwidth]{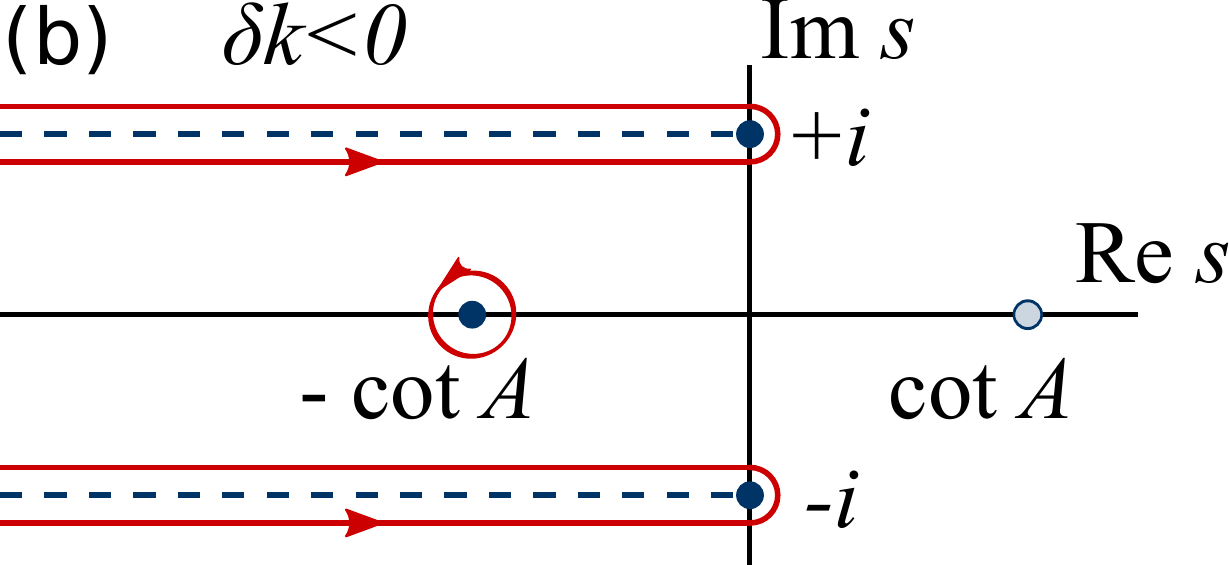}
\caption{Branch cuts (dashed lines) and poles (dots) of the inverse Laplace transform for (a) positive $\delta k$ and (b) negative $\delta k$. By writing ${\rm arccot}\, s=(1/2i)\ln(1+s)/(1-s)$ we have chosen the branch cuts along $s=\pm i-\sigma$ with $\sigma\in[0,\infty)$. With this choice there are no poles for $\delta k>0$ (and $\rho=0$). In the case $\delta k<0$ there is a pole at $-\cot A<0$. The pole in the positive real half plane at $+\cot A$ is regularized by a suitable choice of coefficients ${\cal N}_{1/2}$. The red lines with arrows denote the contour of integration in the inverse Laplace transform (\ref{inverseLaplace_def}).}
\label{fig:branch_cuts}
\end{figure}

Finally we consider all terms in Eq.~(\ref{eigenprob_ndk}) wich are linear in $\delta k$. This yields 
\begin{align}
 \delta M v+ |\delta k|M\delta v+|\delta k|\delta M\delta v=|\delta k|\rho u_0,
\end{align}
for which we evaluate each term separately. The first term becomes 
\begin{align}
(\delta M v)_i&=\frac{i\Delta}{k_Fa}\sum_{j=1}^{i-1} \frac{\sin[\delta k (i-j)]}{i-j}\mathcal{N}_1\beta^j .
\end{align}
The sum is dominated by the first few terms due to the exponential decay $\beta^j$ and since $i\to \infty$, we can safely set $i-j\to i$. Thus, the continuum limit yields 
\begin{align}
(\delta M v)_i\to \delta k\frac{i \Delta}{k_F}\mathcal{N}_1\frac{\beta}{1-\beta}\frac{\sin y}{y} . 
\end{align}
The continuum limit of the second and third term can be obtained in a similar manner as for Eq.~(\ref{deltak_1/2}), 
\begin{align}
 (M\delta v)_i&\to \frac{i\Delta}{k_Fa}A \delta v(y)\\
(\delta M\delta v)_i&\to {\rm sgn}\, \delta k\frac{i\Delta}{k_Fa}\int_0^y dz\frac{\sin(y-z)}{y-z}\delta v(z) .
\end{align}
Thus, we ultimately obtain the integral equation
\begin{align}
&A\delta v(y)+{\rm sgn}(\delta k)\int_0^y dz\frac{\sin(y-z)}{y-z}\delta v(z)\nonumber\\
&=- \frac{i\mathcal{N}_2\rho k_Fa}{\Delta}e^{-\cot(A)y}-{\rm sgn}\, \delta k\mathcal{N}_1a\frac{\beta}{1-\beta}\frac{\sin y}{y}.\label{inteq_ndk}
\end{align}

For the case $\delta k>0$ and with $\rho=0$, this equation reduces to Eq.~(\ref{IntEq}) of the main text. For both signs of $\delta k$, this integral equation (\ref{inteq_ndk}) can be solved by Laplace transformation. Using $\mathcal{L}[ e^{-\lambda y}]=1/(s+\lambda)$ and $\mathcal{L} [\sin y/y]={\rm arccot}\, s$ we find the solution of the integral equation to be
\begin{align}
&\delta v(y)=\mathcal{L}^{-1}\biggl[\frac{-1}{A+{\rm sgn}\, \delta k\, {\rm arccot}\,  s}\nonumber\\
&\left({\rm sgn}\, \delta k\frac{\mathcal{N}_1a\beta}{1-\beta}\,{\rm arccot}\, s +\frac{i \mathcal{N}_2\rho k_Fa}{\Delta}\frac{1}{s+\cot A}\right)\biggr].\label{inverseLaplace_ndk}
\end{align}
\begin{figure}[tb]
\centering
\includegraphics[width=.45\textwidth]{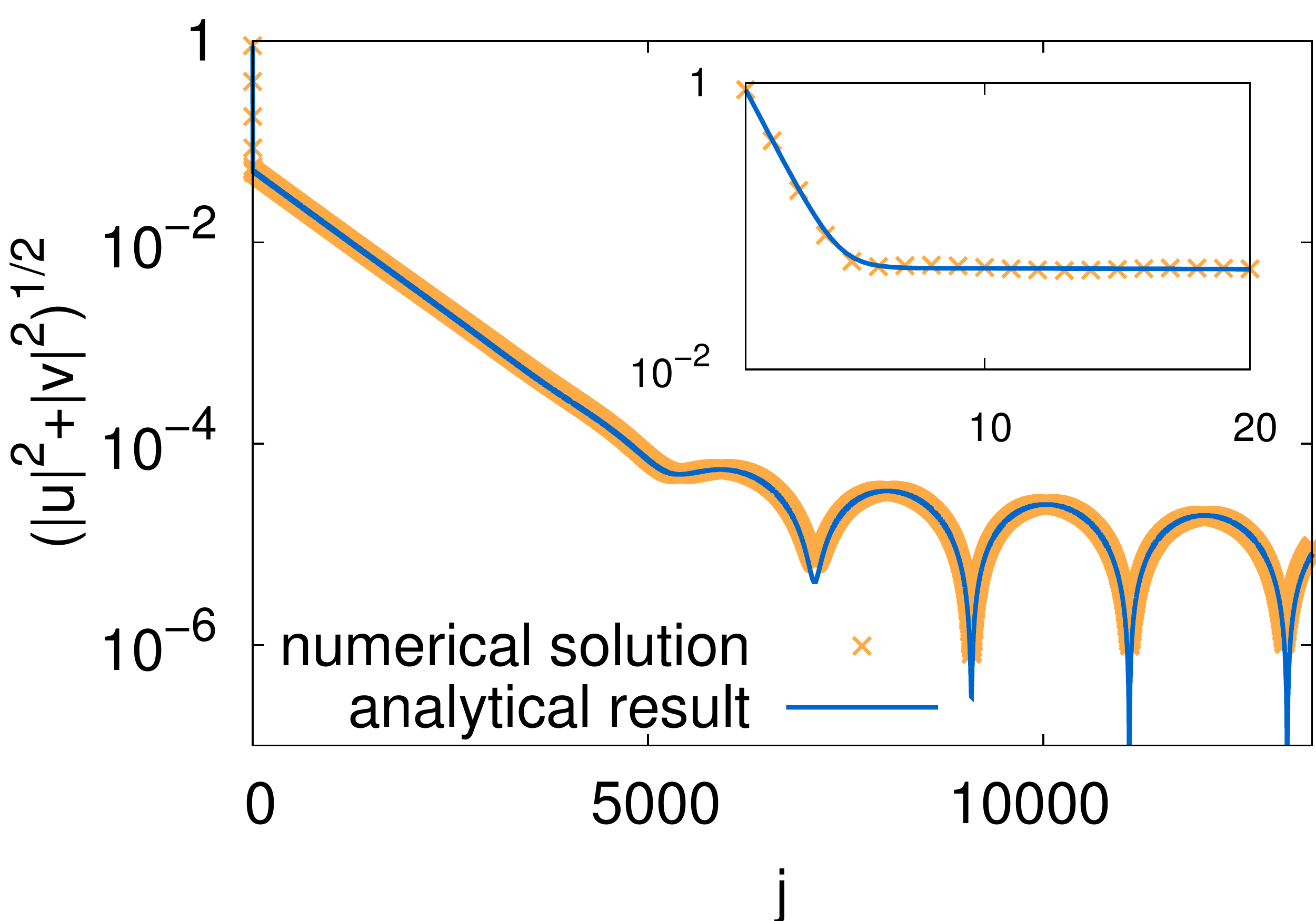}
\caption{Wavefunction modulus $(|u|^2+|v|^2)^{1/2}$ in the nontopological phase ($\delta k<0$) exhibiting a two-exponential decay with a power-law tail. The numerical results from exact diagonalization of the tight-binding BdG Hamiltonian (orange crosses) agree well with the analytical solution $\bigl(\bigl|v_0+|\delta k| \delta v\bigr|^2+|\delta k|\cdot|u_0|^2\bigr)^{1/2}$ (blue line), where we obtained $\delta v$ from the inverse Laplace transform (\ref{inverseLaplace_ndk}) by numerically integrating along the branch cuts in Fig.~\ref{fig:branch_cuts}(b). For the plot we have chosen $\delta k=-5\times 10^{-4} \pi$.}
\label{fig:wv-inteq-ndk}
\end{figure}
The inverse Laplace transform is defined as
\begin{align}
 \mathcal{L}^{-1}f(s)=\frac{1}{2\pi i}\int_{\lambda -i\infty}^{\lambda+i\infty}e^{sy}dsf(s),\label{inverseLaplace_def}
\end{align}
where $\lambda$ is a real number exceeding the real part of all singularities of $f$. For $\delta k>0$, all branch cuts and poles of the Laplace transform in Eq.\ (\ref{inverseLaplace_ndk}) can be chosen to lie in the negative real half plane (see Fig.~\ref{fig:branch_cuts}) and thus the weight $e^{sy}$ is decaying for all $y>0$. In contrast for $\delta k<0$ the Laplace transform has a pole in the positive real half plane at $s_p=\cot A$, which gives rise to an exponentially growing solution in $\delta v(y)$. In order for the wavefunction to be localized at the left end of the chain, we thus have to fix the ratio ${\cal N}_{1}/{\cal N}_2$ in a way that regularizes the Laplace transform at $s_p$, i.e.,
\begin{align}
 \frac{\mathcal{N}_2}{\mathcal{N}_1}=-2iA\cot A\frac{\Delta}{\rho k_F}\frac{\beta}{1-\beta}
\end{align}
for which the numerator in (\ref{inverseLaplace_ndk}) vanishes at $s_p$. Thus by choosing a proper basis in the degenerate subspace, we have ensured that the wavefunction is localized at the left end.  The coefficients  $\mathcal{N}_{1/2}$ are now readily determined in terms of $\rho$ from the overall normalization of the wavefunction $\sum_i(|v_{0,i}|^2+|u_{0,i}|^2)=1$ which is true in the limit $\delta k\to 0$. 

Finally we obtain $\delta v(x)$ from the inverse Laplace transform by integrating along the contour shown in Fig.~\ref{fig:branch_cuts}. The integral can be solved by expanding in $y\gg 1$, 
\begin{align}
\delta v(y)\to \begin{cases}
\displaystyle
4aA\, {\rm sgn}\, \beta\sqrt{\frac{1+\beta}{1-\beta}} \frac{\sin y}{y\ln^2y},&\quad  \delta k>0\\
\displaystyle
4aA\mathcal{N}_1\frac{\beta}{1-\beta} \frac{\sin (y+2A)}{y\ln^2y},&\quad \delta k<0
\end{cases}
\end{align}
and neglecting terms of order $1/y\ln^3y$.

In summary, we have found analytical expressions for the Nambu wavefunction of the subgap states close to the phase transition.
For $\delta k<0$, the bound state at one end of the chain has the form  
\begin{align}
\psi_i= \binom{\sqrt{|\delta k|}u_0\left(i|\delta k|\right)}{v_{0,i}+|\delta k|\delta v(i|\delta k|)}
\end{align}
which comprises the leading order terms in $|\delta k|$ for all sites. On the first few sites, $\psi$ decays exponentially, then crosses over to a much slower exponential decay with amplitude $O(\sqrt{|\delta k|})$ and a decay length $O(1/|\delta k|)$. Further inside the chain the wavefunction decays as a power law with logarithmic corrections. For $\delta k>0$, the wavefunction has a similar form with $u_0=0$.

In Fig.~\ref{fig:wv-inteq-ndk} we compare the analytical solution to the tight-binding wavefunction obtained by exact diagonalization and find excellent agreement.

\subsection*{Localization of Majorana bound states for finite $\xi_0$}

In this section, we briefly discuss in general how the localization of Majorana bound states is affected by a large but finite coherence length $\xi_0$ that limits the range of the hopping and pairing terms. In this case, the topological phase transition is shifted away from the Bragg point [see gray line in Fig.~\ref{fig:Beff}(a) of the main text], but Majorana bound states that are exponentially localized on short scales persist as long the Bragg point is inside the topological phase. In the dispersions $h_k$ shown in the insets of Fig.~\ref{fig:Beff}(b) and (c) of the main text, a finite $\xi_0$ smoothens the steps on the scale $1/\xi_0$. As a consequence the gap closes continuously as a function of $k_F$ in the immediate vicinity of the critical point and we can define an induced coherence length $\xi_{\rm ind}$ that diverges at the phase transition. Deep in the topological phase $\xi_{\rm ind}\sim\xi_0$ (see Ref.~\onlinecite{pientka13} of the main text for details). Thus a finite coherence length does not change the decay properties of Majorana bound states on scales shorter than $\xi_0$, i.e., both the exponential decay with a localization length comparable to the lattice spacing and the power-law tail persist. However, at larger distances the decay becomes exponential again on the scale of the induced coherence length, $\sim \exp(-ja/\xi_{\rm ind})$.

Right at the Bragg point, the tail is always eliminated by destructive Bragg interference and only the short exponential decay survives. Since finite $\xi_0$ reduces the extent of the Shiba wavefunction, this interference becomes less effective and the localization length increases compared to $\xi_0=\infty$ [cf.\ inset of Fig.~\ref{fig:wv}(b) of the main text]. In the vicinity of the Bragg point, the wavefunction acquires a small-amplitude tail as discussed in the previous paragraph, independent of the sign of $\delta k$. Since the power-law tail $\sim 1/y\ln^2y$ sets in only at rather large $y$, the tail is often completely dominated by the slow exponential $\sim \exp(-ja/\xi_{\rm ind})$.


\begin{thebibliography}{99}

\bibitem{review1} J.\ Alicea, Rep.\ Prog.\ Phys.\ {\bf 75}, 076501 (2012).

\bibitem{review2} C.W.J.\ Beenakker, Annu.\ Rev.\ Con.\ Mat.\ Phys.\ {\bf 4}, 113 (2013).

\bibitem{kitaev2} A.Y.\ Kitaev, Phys.\ Usp.\ {\bf 44}, 131 (2001).

\bibitem{fu08} L.\ Fu and C.L.\ Kane, Phys.\ Rev.\ Lett.\ {\bf 100}, 096407 (2008).

\bibitem{lutchyn10} R.M.\ Lutchyn, J.D.\ Sau, and S.\ Das Sarma, Phys.\ Rev.\ Lett.\ {\bf 105}, 077001 (2010).

\bibitem{oreg10} Y.\ Oreg, G.\ Refael, and F.\ von Oppen, Phys.\ Rev.\ Lett.\ {\bf 105}, 177002 (2010).

\bibitem{alicea11} J.\ Alicea, Y.\ Oreg, G.\ Refael, F.\ von Oppen, and M.P.A.\ Fisher, Nature Phys.\ {\bf 7}, 412 (2011).

\bibitem{mourik12} V.\ Mourik, K.\ Zuo, S.M.\ Frolov, S.R.\ Plissard, E.P.A.M.\ Bakkers, and L.P.\ Kouwenhoven, Science {\bf 336}, 1003 (2012).

\bibitem{das12} A.\ Das, Y.\ Ronen, Y.\ Most, Y.\ Oreg, M.\ Heiblum, and H.\ Shtrikman, Nature Phys.\ {\bf 8}, 887 (2012).

\bibitem{bernevig} S.\ Nadj-Perge, I.K.\ Drozdov, B.A.\ Bernevig, and A.\ Yazdani, Phys.\ Rev.\ B {\bf 88}, 020407(R) (2013).

\bibitem{nagaosa} S.\ Nakosai, Y.\ Tanaka, and N.\ Nagaosa, Phys.\ Rev.\ B {\bf 88}, 180503(R) (2013).

\bibitem{loss} J.\ Klinovaja, P.\ Stano, A.\ Yazdani, and D.\ Loss, Phys.\ Rev.\ Lett.\ {\bf 111}, 186805 (2013).

\bibitem{braunecker} B.\ Braunecker and P.\ Simon, Phys.\ Rev.\ Lett.\ {\bf 111}, 147202 (2013).

\bibitem{franz} M.M.\ Vazifeh and M.\ Franz, Phys.\ Rev.\ Lett.\ {\bf 111}, 206802 (2013).

\bibitem{pientka13} F.\ Pientka, L.I.\ Glazman, and F.\ von Oppen, Phys.\ Rev.\ B {\bf 88}, 155420 (2013).

\bibitem{ojanen} K.\ P\"oyh\"onen, A.\ Weststr\"om, J.\ R\"ontynen, and T.\ Ojanen, Phys.\ Rev.\ B {\bf 89}, 115109 (2014).

\bibitem{beenakker_magnetic} T.-P.\ Choy, J.M.\ Edge, A.R.\ Akhmerov, and C.W.J.\ Beenakker, Phys.\ Rev.\ B {\bf 84}, 195442 (2011). 

\bibitem{flensberg} M.\ Kjaergaard, K.\ W\"olms, and K.\ Flensberg, Phys.\ Rev.\ B {\bf 85}, 020503 (2012). 

\bibitem{morpurgo} I.\ Martin, and A.F.\ Morpurgo, Phys.\ Rev.\ B {\bf 85}, 144505 (2012). 

\bibitem{yu} L.\ Yu, Acta Phys.\ Sin.\ {\bf 21}, 75 (1965).

\bibitem{shiba} H.\ Shiba, Prog.\ Theor.\ Phys.\ {\bf 40}, 435 (1968).

\bibitem{rusinov} A.I.\ Rusinov, Zh.\ Eksp.\ Teor.\ Fiz.\ Pis'ma Red.\ {\bf 9}, 146 (1968) [JETP Lett.\ {\bf 9}, 85 (1969)].

\bibitem{rmp} A.V.\ Balatsky, I.\ Vekhter, and J.-X.\ Zhu, Rev.\ Mod.\ Phys.\ {\bf 78}, 373 (2006). 

\bibitem{hasan} M.Z.\ Hasan and C.L.\ Kane, Rev.\ Mod.\ Phys.\ {\bf 82}, 3045 (2010).

\bibitem{foot1} Additional Bragg points exist for $2k_F=\pm 2k_h+2\pi n$.

\bibitem{supp} See Supplementary Material at http://link.aps.org/ where we discuss the two-channel phase and the case of finite $\xi_0$ and provide a detailed derivation of the Majorana wavefunction at and near the Bragg point.

\bibitem{simon08} P.\ Simon, B.\ Braunecker, and D. Loss  Phys.\ Rev.\ B {\bf 77}, 045108 (2008).

\bibitem{braunecker09} B.\ Braunecker, P.\ Simon, and D. Loss  Phys.\ Rev.\ B {\bf 80}, 165119 (2009).

\bibitem{Zindex} S.\ Ryu, A.P.\ Schnyder, A.\ Furusaki, and A.W.W.\ Ludwig, New J.\ Phys.\ {\bf 12}, 065010 (2010).

\bibitem{foot2} Incidentally, the two topological phases in Fig.~\ref{fig:Beff}(a) differ by the winding direction and have topological indices $\pm 1$. Thus, the intermediate phase transition involves two simultaneous gap closings.

\bibitem{Rotation} I.\ Adagideli, P.M.\ Goldbart, A.\ Shnirman, and A.\ Yazdani, Phys.\ Rev.\ Lett.\ {\bf 83}, 5571 (1999).

\bibitem{foot3} The second positive-energy subgap state is localized at the right end with electron and hole components exchanged. The two negative-energy states are related to the positive ones by chiral symmetry.

\end{thebibliography}

\begin{thebibliography}{12}

\bibitem{polyanin} A.D.\ Polyanin, A.V.\ Manzhirov, {\em Handbook of integral equations} (Chapman \& Hall/CRC Press, Boca Raton, 2012).

\end{thebibliography}
\end{document}